# Understanding the temperature conditions for controlled splicing between silica and fluoride fibers

Antreas Theodosiou*, *Member, IEEE*, Ori Henderson-Sapir*, *Member, IEEE*, Yauhen Baravets, Oliver T. Cobcroft, Samuel M. Sentschuk, Jack A. Stone, David J. Ottaway, and Pavel Peterka, *Member, IEEE*

*Abstract*—This study explores the efficacy of thermal splicing conditions between silica and zirconium-fluoride fibers, focusing on achieving mechanical strength between the two fibers. A comprehensive characterization of the thermal profile in the hot zone of the filament splicer was conducted using a fiber Bragg grating, providing valuable insights into its stability and overall performance. Results demonstrate mechanically strong joints and suggest a very narrow temperature window to achieve strong connection between the two materials. Moreover, we characterize the surface composition of the $ZrF_4$ fiber using energy dispersive spectroscopy following splicing at ideal temperatures, as well as at higher and lower temperatures. This work paves the way towards future implementation of silica and fluoride fibers splicing using alternative splicing solutions such as $CO_2$ laser system while raising interesting facts for further studies in the specific field.

## I. Introduction

MID-infrared (mid-IR) fiber lasers have emerged as pivotal components in optical sciences, offering distinctive capabilities with applications spanning diverse scientific and industrial domains. Operating within the wavelength range of approximately 2 to 18 micrometers [1], mid-IR lasers exhibit characteristics that render them indispensable for addressing specific challenges across various fields. The intrinsic importance of mid-IR fiber lasers is particularly evident in chemical sensing and spectroscopy applications. Their wavelength range aligns strategically with crucial molecular absorption bands, enabling the identification and analysis of chemical compounds in gases and liquids. Consequently, mid-IR lasers find applications in environmental monitoring [2], [3], gas sensing [4], [5], and medical diagnostics [6], [7].

In the medical domain, mid-IR lasers are crucial in minimally invasive surgical procedures and medical imaging. Their efficiency in interacting with water and biological tissues positions them as valuable tools for cutting, ablation, and imaging in medical interventions [8], [9]. Extending beyond medical applications, mid-IR lasers are critical in material processing, providing precision in working with materials like polymers and ceramics. Their distinctive wavelength characteristics make them suitable for applications in free-space optical communication, where their performance is less susceptible to atmospheric absorption and scattering effects [10], [11]. Within the defense and security realm, mid-IR lasers contribute to infrared countermeasure systems, safeguarding against heat-seeking missiles [10], [12].

Fiber laser configurations offer significant advantages compared with other types of solid-state lasers; therefore, various fiber laser schemes have been developed that minimize free space components and sometimes achieve completely monolithic, all-fiber designs. Unfortunately, the well-established silica fiber technology does not support wavelengths longer than ~2.1 µm. As a result, different types of soft-glasses are used, mostly fluoride materials such as ZBLAN and Indium fluoride, which can expand the transmission window up to 5 µm [13].

Various mid-IR fiber laser demonstrations used doped active fluoride optical fibers as the gain medium while the rest of the laser resonators were assembled using free space components [14]–[16]. To develop the most robust mid-IR fiber lasers, it is essential to avoid using free space elements and replace them with optical fibers. The pumping sources which often operate in the near infrared (NIR) or visible wavelength range and delivered using silica fibres need to be coupled into the active fluoride fiber, and as a result, low loss and strong splicing joints between the silica and fluoride fiber are highly desirable.

ZBLAN and Silica are very different glasses, especially when considering their melting temperature. Silica has a melting temperature of ~2100 °C awhile ZBLAN melts at ~460°C [17]–[19]. Several techniques have been tried to join fibers made from the two materials. The first attempts at splicing the two fibers used arc splicing [20], which did not allow sufficient temperature control at low ranges, even with very low currents. The temperature was unstable, and consequently, the success rate was very low. In 2014, thermal splicing was proposed as an alternative solution for splicing soft glass fibers, and to our best knowledge, the work of Al-Mahrous et al. [21] was the first successful attempt to splice silica and ZBLAN fibers thermally. In this work, a thin Kanthal wire band was folded on an "M-shape" as the heating element where the fibers were placed in the center. However, the losses of these splicing joints varied significantly while the joint strength was extremely weak.

* These authors contributed equally.
A. Theodosiou, Y. Baravets, and P. Peterka are with the Institute of Photonics and Electronics, The Czech Academy of Sciences, Chaberska 57, 182 00 Prague 8 – Kobylisy, Czech Republic.
O. Henderson-Sapir, O.T. Cobcroft, S.m. Sentschuk, J.A. Stone, and D.J. Ottaway, are with the Institute for Photonics and Advanced Sensing, School of Physics, Chemistry and Earth Sciences, The University of Adelaide, Adelaide, 5005, SA, Australia, and with The Australian Research Council, Centre of Excellence for Gravitational Wave Discovery (OzGrav). (e-mail: ori.henderson-sapir@adelaide.edu.au).

In 2015, Zhi-Jian et al. [22] used a commercial filament splicer, Vytran GPX-3000, in which, the working principles were similar to [21] but more controllable. The reproducibility of the method was improved, and on average, a loss of 0.8 dB was achieved, significantly better compared with [21]. The authors demonstrated the first thermal splicing between silica and ZBLAN fiber used in a high-power laser regime, while a year later [23], they successfully incorporated the splicing joints into a supercontinuum mid-IR laser [23]. However, there is no mention of the splicing strength. In 2016, another work from Huang et al. [24] investigated theoretically how to optimize splicing parameters, such as splice offset relative to center of the filament and heating time, by studying how the temperature is distributed along the two fibers with respect to different splicing offsets. Guided by these simulation results, they achieved optimized splicing joints with mean losses of 0.37 dB and a minimum loss of 0.1 dB. No information was provided regarding splice joint strength.

In 2019, Cozic et al. [25] presented thermal splicing joints between a multimode 200 μm core silica fiber and a double clad, single mode fluoride fiber with excellent optical losses and mechanical strengths >300 g, while the authors claim there are no issues with the reproducibility of the method. It is important to note that after the splicing, the authors coated the splice with a low refractive index UV curable resin and then performed the strength measurement [22]. Finally, in 2021, Yang et al. [26] presented the most comprehensive study on this topic, where they experimentally studied and characterized the splice strength and loss between a single mode fluoride and a silica fiber by adjusting the filament power and splice offset. They achieved >300 g and splicing losses <0.4 dB. Table 1 summarizes the specifications of the fibers used in the previous works, including core and cladding diameters and NA.

Beyond thermal splicing, a few other techniques have been proposed, such as mechanical splicing [27], glue splicing [28], [29], using UV resins to enhance the strength of the splice [25] and some additional attempts of using arc fusion splicing [30], [31]. Finally, although recent works have proposed splicing using a $CO_2$ laser fiber processing unit [32], Rivoallan and Guiloux first used this method in 1988 by [33]. With all these trials, filament thermal splicing appears to be the most reliable and reproducible method of splicing and handling fluoride and soft-glass fibers in general.

Following the results in the literature over the last few years on the specific topic, we can note the progress in the loss and strength of the splicing joints with more focus in the splicing loss. However, additional measurements are necessary to reproduce the results from the literature and obtain a better understanding of the mechanism behind the splicing of these two highly dissimilar materials. Fiber pairs spliced in previous works are shown in Table 1. Commonly missing information in all previous works is filament characteristics used for the splicing, such as type and diameter, while splicing parameters other than power and splicing offset were not mentioned.

This work focuses on the mechanical properties of the splices while it emphasizes the correct splicing conditions between the Zirconium-Fluoride ($ZrF_4$) and silica fiber in terms of temperature using the Vytran Splicer GPX-3400 and a graphite filament FTAV2 provided by Thorlabs. Previous studies have primarily addressed fiber splice loss; hence, this work concentrates on the mechanical strength of the splicing. We focus on the calibrating of all the important parameters of the Vytran system, such as power, argon flow, and offset splicing, in terms of temperature. In this way, we create a material guideline map that can be followed and applied by any fusion splicing system, assuming it can maintain high accuracy of appropriately low-temperature splicing conditions.

**Table 1.** Specifications of the silica and fluoride fibers used in previous works, NA – numerical aperture.

| Silica fiber | | | Fluoride fiber | | | Splicing strength | Ref |
|---|---|---|---|---|---|---|---|
| Core (μm) | Clad. (μm) | NA | Core (μm)) | Clad. (μm) | NA | | |
| 3.89 | 125 | 0.13 | 2.4 | - | 0.25 | - | [21] |
| 9 | 125 | 0.14 | 9 | 123 | 0.2 | - | [22] |
| 7 | 125 | 0.2 | 9 | 125 | 0.2 | - | [23] |
| 3.4 | 125 | 0.22 | 2.7 | 123 | 0.25 | - | [24] |
| 200 | 200 | 0.22 | 14 | 250/290 | - | >300 g | [25] |
| 7 | 125 | 0.2 | 7.5 | 148 | 0.23 | >300 g | [26] |

## II. System Calibration

To better understand the thermal splicing procedure using the Vytran GPX3400 system, we calibrated the system's parameters including power, argon flow, and distance from the filament center in terms of temperature. To do that, we used a commercial femtosecond laser inscribed fiber Bragg grating from Lumoscribe Ltd [34] with a total length of ~2 mm as the temperature sensing element that was interrogated by a broadband NIR light source developed in-house at the Institute of Photonics and Electronics. The reflection spectrum of the FBG was monitored using an IBSEN IMON USB256 spectrometer with wavelength resolution of 169 pm. The data were post-processed using the Zero-Crossing Algorithm to optimize the resolution of our FBG measurements [35]. Our experiments were performed using the graphite filament FTAV2 from Thorlabs for all the results presented in this study. In all calibration experiments, the center of the FBG was placed at the geometric center of the Vytran's Ω shaped filament before the temperature measurements. The alignment was performed using the Vytran's microscope viewing system, illuminated by an external light source.

We began by calibrating the power of the system, starting with very low power set values of 0.3, 0.5, 0.7, and 1 W, and then proceeded to higher powers of 3 and 5 W. The filament was set to be active for 10 seconds, and all the other parameters were deactivated including filament ramp. The reference value for argon flow was set at 0.3 l/min, with filament power the only variable. Figure 1 presents the FBG's wavelength response with respect to the time when the filament was activated for 10 seconds while operating at power levels of 1, 3, and 5 W. It is important to note that these power values corresponded to the set power in the Vytran's user interface that could vary from the actual power values. The actual power dissipated by the filament was based on a power offset value calibrated daily which is generally affected by ambient temperature, humidity, air pressure and filament aging.

The temperature coefficient of an FBG in silica fiber is 10.22 pm/°C [36]. According to this value, we proceeded to translate the wavelength shift of the FBG in terms of temperature, as presented in Fig. 2. We noticed a linear response between temperature and the Vytran's power reading with a slope of 55 °C/W. Moreover, to examine the power stability of the system, we performed multiple trials at 1 W while monitoring the response of the FBG. Our results, presented in Figure 3, demonstrate a temperature stability of <0.25°C of the system. As can be observed in Figures 1 and 3, the temperature change in the filament, as measured using the FBG, is not a flat step change but shows a slowly decreasing positive gradient until it reaches the highest value. This temperature difference between the start and stop of the gradient is around 4°C. We used the highest temperature value recorded for our calculations to ensure accuracy and consistency in our data analysis.

In the same manner, we repeated our thermal calibration experiments, but this time, kept the power constant at 1 W while changing the Argon flow by steps of 0.05 l/min. The results can be seen in Figure 4. According to our measurements, the slope between argon flow and temperature was found to be -12.6 °C per 0.1 l/min flow; so, the higher the argon flow, the lower the temperature.

Finally, in Figure 5, we can see the temperature distribution of the specific filament for 1, 2, and 3 W of set power when the FBG was placed with an offset from the center of 600, 1200, 1800, 2400, and 3600 μm.

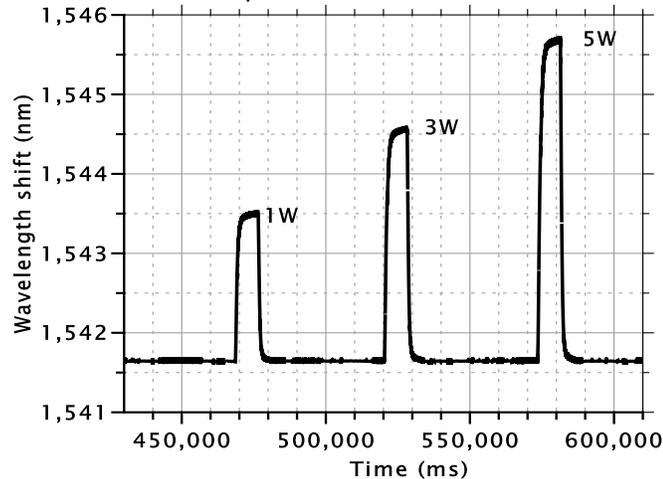

**Figure 1:** Time dependent wavelength shift of the fiber Bragg when subjected to 1, 3, and 5 W of filament power.

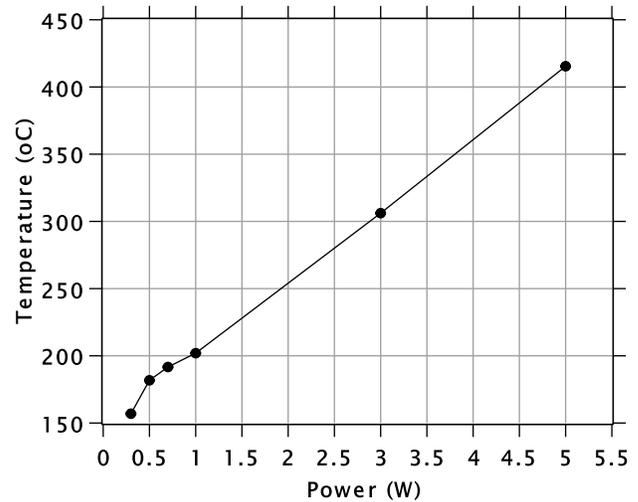

**Figure 2:** Filament temperature variation with respect to filament power.

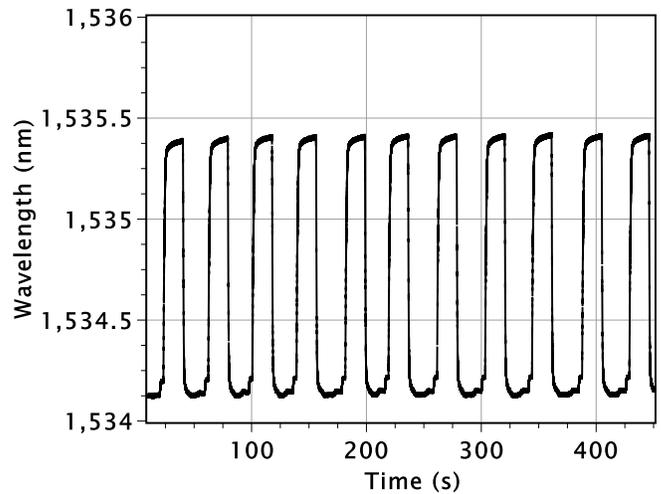

**Figure 3:** Power stability measurements of the splicing filament as measured using an FBG when repeating the same filament heating profile.

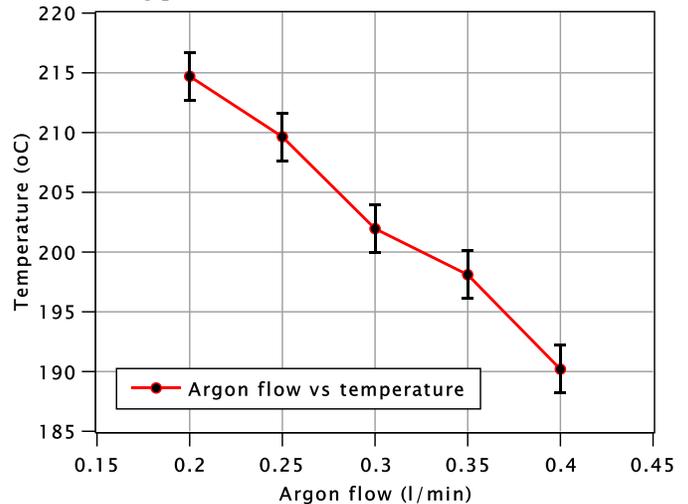

**Figure 4:** Temperature dependence of the filament with respect to the argon flow. The slope was found to be -12.6 °C per 0.1 l/min.

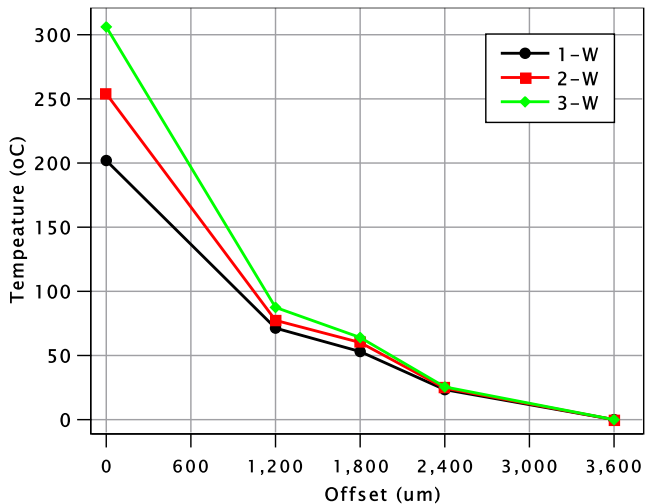

**Figure 5:** Temperature dependence at the center of the FBG with respect to the filament offset position.

### III. Splicing Methodology and Results

Following the previous characterization of the filament temperatures, we can now choose the correct parameters for splicing between the silica and ZrF$_4$ fibers. For our experiments, a double cladding ZrF$_4$ fiber was used which was manufactured by Le Verre Fluoré, France. The fiber had a core diameter of 16.5 μm and 260 μm diameter for the first cladding. A low-index polymer with a diameter of 290 μm acts as the second cladding. The numerical aperture (NA) of the core is 0.14 with a cut-off wavelength of 2.7 μm. The core is doped by 10000 ppm of ErF$_3$ ions. In comparison, silica fiber is a common single-mode SMF28 with diameters 8.2 μm and 125 μm for core and cladding, respectively.

To prepare the 260 μm ZrF$_4$ fiber for splicing, the jacket and FEP coating were removed. The jacket was carefully scored using a scalpel while the fiber was secured to a plate. The fiber was then cleaned off using methanol-soaked wipe and the jacket gently removed using a scalpel.

Our strategy is to have an offset of the filament hot zone towards the silica fiber. Heat the silica fiber sufficiently before touching the two fibers together to ensure uniform heating along and across the silica fiber, and then quickly join the two fibers. At the same, it is important to avoid large offsets, because they require higher temperatures which have a higher possibility of melting the fiber at the heating point. In addition, higher temperatures result in more significant sagging and bending of the silica fiber, which increases the splice losses.

On the other hand, since the temperature decreases along the length of the fiber, it is necessary to heat the fiber for sufficient time to ensure appropriate temperature at the facet of the SMF28 fiber [37]. In our case, we choose 5 seconds as a pre-splicing heating time (Hot push delay). Experimentally, we found that is possible to achieve splicing for pre-splicing heating time higher than 3 seconds. To the authors' best knowledge, there is no information in the literature about the pre-splicing heating preparation of the silica fiber. The power and argon flow were the only two parameters adjusted to optimize the quality of the splicing joints. Based on the temperature calibration of the system presented in section II of this paper and the 0.1 W of power and 0.01 l/min argon flow resolution available with the GPX-3400 system, we can control the temperature at the center of the filament with an accuracy of ~0.6 °C considering the power of the filament as the primary mechanism to set the splicing temperature and the argon flow for fine tuning. The splicing duration was set at 0.2 seconds, and the speed of the hot push was set at 175 μm/s to obtain a hot push distance of 35 μm. The 350 μm offset of the filament results in a 21.3% lower temperature at the splice joint according to the results presented in Figure 5.

To determine the optimal temperature for splicing these two fibers, we adjusted the filament power and characterized the mechanical strength of the joint immediately after splicing. For the mechanical strength characterization, we employed the tension monitor of the Vytran system and applied axial tension to the joints using the fiber holders of the Vytran.

The initial characterization of the splicing joint strength at varying temperatures was conducted by only using power adjustment, as noted by the blue circles in Figure 6. The power resolution available on the Vytran was 0.1 W which corresponds to 5.5 °C per step. After noting the mechanical strength of the joint for power values 9.0, 9.1, and 9.2 W, we optimize the splicing temperature by adjusting the argon flow by 0.01 l/min. As mentioned previously, this adjustment results in a temperature adjustment of 0.6 °C, filling the gaps in Figure 6.

Figure 6 shows that there is a narrow temperature window of 7 °C (noted with a red rectangle in Figure 6) to achieve an acceptable mechanical strength splice joint > 60 gr, and an even narrower window of 2.5 °C where more mechanically robust splice joints exhibiting >150 gr of proof-testing are achieved (red hashed area).

Similar but slightly weaker joints were observed when splicing passive, double clad, Le Verre Fluoré fibre with 14 μm core and 260 μm cladding diameters, respectively. This investigation was conducted using an older Vytran system, potentially explaining the weaker joints due to poorer current stability.

Subsequently, each time the filament is normalized, it is highly important to consider the normalization power offset in calculations with the highest possible accuracy. Moreover, it is imperative for the fiber to be well-aligned laterally in the center of the filament, otherwise, the temperature may vary. The temperature at the offset position of the silica fiber was selected to be 452.2 °C, which resulted in 355.9 °C close to the end face of the ZrF$_4$ fiber.

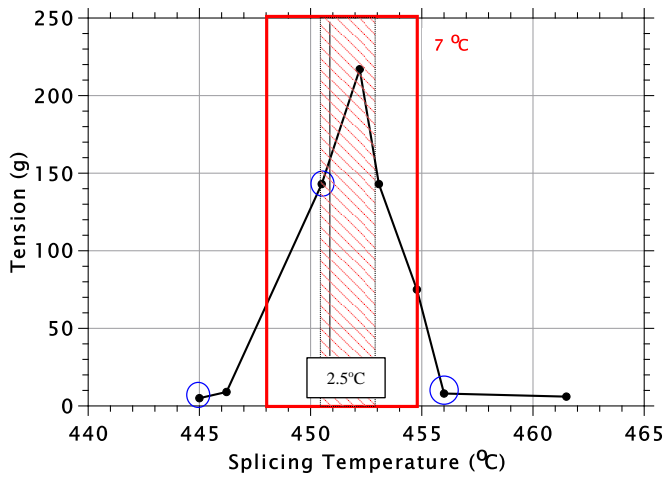

**Figure 6:** Mechanical strength of the splice joints with respect to temperature at the center of the filament's hot zone. The splices were performed with a filament offset of 350 μm towards the silica fiber.

According to the literature, the glass transition temperature of ZrF$_4$ is ~260 °C, and the melting point temperature is > 450 °C [17]. A microscope picture of a splicing joint with a mechanical strength of 225 g spliced with a temperature of 452.2 °C at the offset position is presented in Figure 7a, while the 67 g spliced with a temperature of 454.8 °C is presented in Figure 7b. We can observe the rounding of the ZrF$_4$ fiber edges when the temperature is not ideal for splicing. The splicing parameters used for this study are presented in Table 2. However, various other combinations can be selected to achieve the appropriate splicing temperature, based on the calibration curves mentioned in this work.

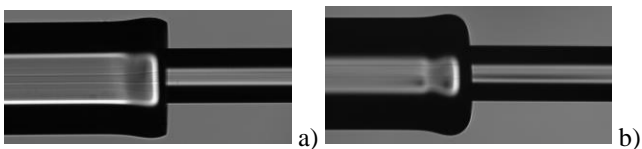

**Figure 7:** Microscope picture of a splicing joint, ZrF$_4$ fibre on he left. Temperature at the offset position of a) at 452.2 °C resulting to >220gr mechanical strength and b) at 454.8 °C resulting to 67 gr mechanical strength.

**Table 2:** Parameters used for optimum splicing between ZrF$_4$ and SMF28 fiber

| Splicing Parameters | |
|---|---|
| Power (W) | 5* |
| Duration (s) | 0.2 |
| Hot Push (μm) | 35 |
| Hot Push Delay (s) | 3-5 |
| Push Velocity (μm /s) | 175 |
| Offset (μm) | 350 |
| Argon flow (l/min) | 0.3** |
| Pre-push gap (μm) | 5 |
| Pre-gap (μm) | 15 |

*Set power, the filament offset power need to be considered.
**The argon flow should be modified accordingly to achieve the required splicing temperature.

An interesting phenomenon was observed in the fibers after tension proof-testing. After the tension test and when the splicing joint was destroyed, we observed a rectangular shape at the end facet of the ZrF$_4$ fibers, see Figure 8. The length of the structure was highly dependent on the splicing temperature. Higher temperatures resulted in a longer structure. Although useful as an indicator for the correct splicing temperature, the exact mechanism behind this behavior is currently not clear. Figures 8a-d present microscope images for splices performed at 452.9 °C, 456.8 °C, 467.0 °C and 467.3 °C, respectively. Harbison showed that ZBLAN fibers require a viscosity of 10$^5$ poise for approximately 2 seconds at the fiber end to achieve effective splicing [38]. This viscosity can be achieved at a temperature of approximately 340 °C. However, this temperature can vary depending on the fiber's composition. Based on our results, we can conclude that there is only a small window of ~2.5 °C to reach the appropriate viscosity and it is shifted by ~ 11 °C when compared to the [38], likely due to the different composition of the fibers used in this work compared with standard ZBLAN composition that was used by Harbison.

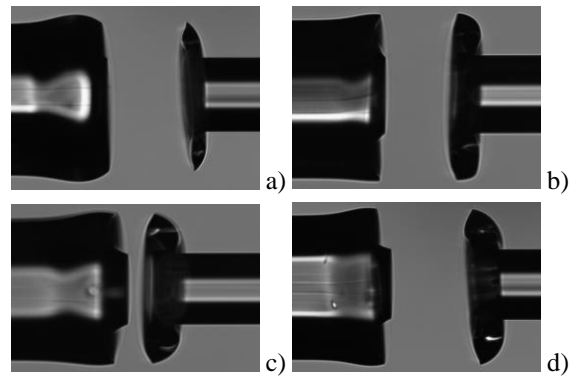

**Figure 8:** Microscope images of splices performed at different temperatures, ZrF$_4$ fibre on he left. a) 452.9 °C b) 456.8 °C c) 467.0 °C d) 467.3 °C. Considering the optimal temperature to be 451.2 °C.

Moreover, a more careful inspection of the end face of the structure as presented in Figure 9, shows that the longer structure was also associated with the appearance of bubble-like shapes on the end face of the ZrF$_4$ fiber. The number of bubbles increase while their size decreases with temperature increasing further from the optimal splicing temperature allowing for a practical way to determine whether the splicing temperature is too high. At splicing temperatures lower than the optimal, pitting and some bubbles appeared, but they tended to be much less numerous and irregular in shape. Figure 9a-c presents the optical microscope images of the ZrF$_4$ fiber end faces when the optimal splicing temperature is achieved (9a), when a lower temperature than the optimal is achieved (9b), and when a higher temperature is achieved (9c), respectively.

It is worth noting that a single bubble is always present on the core of the ZrF$_4$ fiber, suggesting a potential significant loss mechanism in coupling between the two fibers. In addition, no features are observed on the silica fiber's end face.

To investigate further this phenomenon, we observed the bubbles using scanning electron microscopes (SEM), Figure 9d.

One equipped with an energy dispersive spectroscopy (EDS) and another with a focused ion beam (FIB). Initially, one can see that there is no sign of crystallization, unlike the cases reported by [33], [39]. In addition, the surface appears to have undergone significant outgassing in the form of small bubbles. The FIB data of the bubbles investigated suggests that the bubbles consist of a void that is covered by a thin glass membrane. From the EDS results, we found that the elemental composition underneath the bubbles have a different composition than the surrounding, undisturbed fiber endface. The bubbles including the smaller ones, both in the core and in the cladding (having slightly different composition to ensure refractive index change), contain ~ 10% more barium (Ba) and equivalently less fluorine (F) than the surrounding undisturbed areas. At high magnification (Figure 9d), the bubble surfaces appear featureless, however, milling with the FIB beam collapsed the bubbles exposing the void underneath their thin membrane. EDS spectra from various sites show consistently higher Ba and lower F percentages in bubbled areas compared to the background material. This suggests that all bubbles, regardless of size, have elevated Ba and reduced F levels.

The elemental changes, surface features and the appropriate temperature range suggest the potential formation of $SiF_4$ gas bubbles during the hot push at temperatures higher than optimal, where the $BaF_2$ from the $ZrF_4$ fiber contacts the silica fiber [40]. The presence of bubbles in the core, even at optimal temperatures, could result from the formation of $GeF_4$ gas bubbles due to the interaction between the Ge-doped core of standard silica fiber (such as in an SMF28) and the BaF from the ZBLAN fiber [40]. Unfortunately, it is not possible to test for the presence $GeF_4$ or $SiF_4$ using the instruments currently available to us, therefore further evaluation of this hypothesis will be the focus of future work.

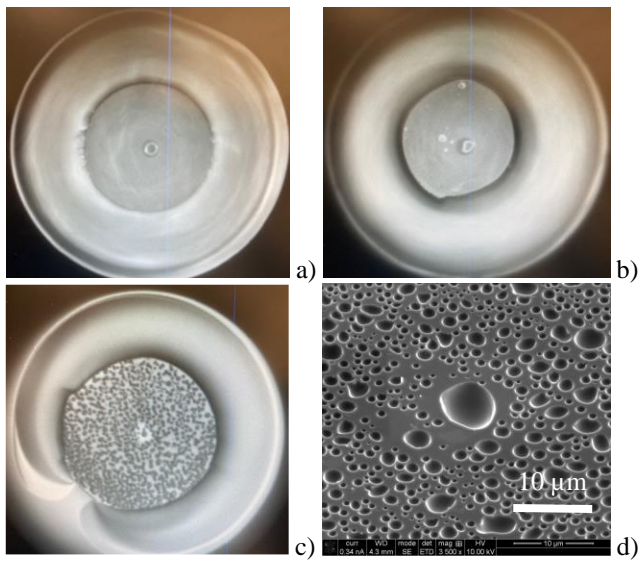

**Figure 9:** Microscope images of the bubbles observed on the ZBLAN passive fiber endface. Note that the inner circle is the area where the silica fiber interfaces with the $ZrF_4$. The external ring is where the $ZrF_4$ fiber encases the silica fiber creating the splice joint. a) Optimal splicing temperature, no bubbles except directly on the core, **b)** splicing temperature lower than the optimal, **c)** Splicing temperature higher than the optimal, **d)** SEM imaging of the bubbles on the endface of the $ZrF_4$ fiber in a similar profile to the one shown in **(c)**.

## IV. CONCLUSION

In this work, we presented a comprehensive study on thermal splicing using a filament splicer to connect fluoride ZBLAN fibers to silica fiber. We characterized the system parameters in terms of temperature and tried to interpret and understand the splicing procedure in terms of temperature and not power, as presented in previous works. We propose to the best of our knowledge for the first time, a methodology and a roadmap for other researchers and professionals to follow, using similar splicing equipment or other methods such as $CO_2$ lasers or electrode techniques. Our measurements show that there is a very small temperature window of <2.5 °C for strong mechanical splices between the two materials. We suggest an alternative method that precisely regulates the splice temperature with a resolution of approximately 0.6°C to achieve these outcomes. We analyze and discuss the composition at the end face of the $ZrF_4$ fiber when subjected to various temperatures and we show that there is an interesting correlation between splicing temperature, the strength of the spliced joints and the concentration of F and Ba elements concentration. Our results indicate that a chemical reaction could be responsible for the formation of bubbles on the end face of the $ZrF_4$ fiber. The presence of bubbles in the core, even at optimal temperatures, could result from the formation of $GeF_4$ gas bubbles due to the interaction between the Ge-doped core of standard silica fiber and the BaF from the ZBLAN fiber. We believe that this work set an important step towards efficient and reproduceable fusion splicing between different optical fiber materials using thermal splicing.

### ACKNOWLEDGMENT

The authors would like to thank Prof. Heike Ebendorff-Heidepriem for the useful discussions and Dr. Ken Neubauer and Dr. Animesh Basak for providing microscopy results.
The work is co-funded by the European Union and the Ministry of Education, Youth and Sports under the Marie Sklodowska-Curie Czech grant No CZ.02.01.01/00/22_010/0001912. This work was performed, in part, at the OptoFab node of the Australian National Fabrication Facility supported by the Commonwealth and SA State Government and an Australian Research Council Discovery Grant 220102516 DP and Centre of Excellence Grant CE170100004. The authors acknowledge the instruments and expertise of Microscopy Australia at Adelaide Microscopy, The University of Adelaide, enabled by NCRIS, university, and state government support.